\newif\iffigs\figstrue
\documentclass[12pt]{article}
\iffigs
   \input{epsf}
\else
\fi

\newcommand{\sect}[1]{\setcounter{equation}{0}\section{#1}}

\newcommand{\app}[1]{\setcounter{section}{0}\setcounter{equation}{0}
\renewcommand{\thesection}{\Alph{section}}
\section{#1}}
\newcommand{\eq}{\begin{equation}}
\newcommand{\eqa}{\begin{eqnarray}}
\newcommand{\en}{\end{equation}}
\newcommand{\ena}{\end{eqnarray}}
\newcommand{\enn}{\nonumber \end{equation}}

\def\sk{\vskip .4cm}
\def\noi{\noindent}
\def\om{\omega}

\def\al{\alpha}
\def\la{\lambda}

\def\ga{\gamma}
\def\Ga{\Gamma}

\def\Cb{\bar{C}}

\def\rhop{{\rho}^{\prime}}

\def\rhopp{\rho ''}

\def\epsi{\varepsilon}
\def\we{\wedge}

\def\de{\delta}

\def\part{\partial}

\def\R#1#2{ R^{#1}_{~~~#2} }

\def\L#1#2{ \La^{#1}_{~~~#2} }

\def\La{\Lambda}

\def\Cb{{\bf \mbox{\boldmath $C$}}}

\def\c#1#2{ C^{#1}_{~~#2} }
\def\C#1#2{ {\bf \mbox{\boldmath $C$}}^{#1}_{~~#2} }

\def\DR{\Delta_R}
\def\DL{\Delta_L}

\def\D{\Delta}

\def\limepsizero{\lim_{\epsi \rightarrow 0}}

\def\n2{{{N+1} \over 2}}

\def\square{{\,\lower0.9pt\vbox{\hrule \hbox{\vrule height 0.2 cm
\hskip 0.2 cm \vrule height 0.2 cm}\hrule}\,}}

\def\Q.E.D.{\rightline{$\Box$}}

\def\sumong{\sum_{g \in G}}
\def\sumongp{\sum_{g \in G'}}
\def\sumongnote{\sum_{g \not= e}}

\def\sumonh{\sum_{h \in G}}
\def\sumonhp{\sum_{h \in G'}}

\def\Lcal{{\cal L}}
\def\Rcal{{\cal R}}

\def\tti{{\tilde t}}
\def\omc#1#2{\om^{#1}_{~~#2}}
\def\W#1#2{W^{#1}_{~~~#2} }
\def\V#1#2{V^{#1}_{~#2} }
\def\Gc#1#2{\Ga^{#1}_{~~#2}}
\def\a#1#2{a^{#1}_{~#2}}
\def\ainv#1#2{(a^{-1})^{#1}_{~#2}}

\begin{document}
\begin{titlepage}
\vskip -1cm
\rightline{DFTT-44/99}
%\rightline{LBNL-40824}
\rightline{October 1999}
\vskip 1em
\begin{center}
{\large\bf Gravity on Finite Groups }
\\[2em]
Leonardo Castellani\\[.4em]
{\sl Dipartimento di Scienze e Tecnologie Avanzate,
 East Piedmont University, Italy; \\
Dipartimento di Fisica Teorica and Istituto Nazionale di
Fisica Nucleare\\
Via P. Giuria 1, 10125 Torino, Italy.} \\
{\small castellani@to.infn.it}\\[2em]
\end{center}
\vskip 4 cm
\begin{abstract}
Gravity theories are constructed on finite groups $G$. A
self-consistent review of the differential calculi on finite $G$
is given, with some new developments. The example of a bicovariant
differential calculus on the nonabelian finite group $S_3$ is
treated in detail, and used to build a gravity-like field theory
on $S_3$.
\end{abstract}
%\noi{April 1998}~~\hskip 10cm~q-alg/9709032
%\centerline{{\small to appear on {\bf European Phys. Jou. C}, Particles and Fields }}
\vskip 8cm
\noi \hrule
\vskip .2cm
 \noi{\centerline {\footnotesize
  Supported in part by   EEC  under TMR contract
 ERBFMRX-CT96-0045}}
\end{titlepage}
\newpage
\setcounter{page}{1}
%%%%%%%%%%%%%%%%%%%%%%%%%%%%%%
\sect{Introduction}

The algebraic treatment of differential calculus in terms of  Hopf
structures allows to extend the usual differential geometric
quantities (connection, curvature, metric, vielbein etc.) to a
variety of interesting spaces that include quantum groups,
noncommutative spacetimes (i.e. quantum cosets), and discrete
spaces.
 \sk
 In this paper we concentrate on a particular sort of
discrete spaces, i.e. finite group ``manifolds". As we will
discuss, these spaces can be visualized as collections of points,
corresponding to the finite group elements, and connected by
oriented links according to the particular differential calculus
we build on them. Although functions $f \in Fun(G)$ on finite
groups $G$ commute, the calculi that are constructed on $Fun(G)$
by algebraic means are in general noncommutative, in the sense
that differentials do not commute with functions, and the exterior
product does not coincide with the usual antisymmetrization of the
tensor product.
 \sk
 The physical motivations for finding
differential calculi on finite groups are at least threefold in
our opinion:
 \sk
  i) the possibility of using finite group spaces
as internal spaces for Kaluza-Klein compactifications of
supergravity or superstring theories. Harmonic analysis on such
spaces is far simpler than on the usual smooth manifolds (coset
spaces, Calabi-Yau spaces, etc.) or orbifolds. We note in this
respect that compactification of $D=5$ Yang-Mills theory on the
finite group space $Z^2$ yields precisely the Higgs potential, and
gives it a geometric {\sl raison d' \^etre} . In fact Connes'
reconstruction of the standard model in terms of noncommutative
geometry \cite{Connes} can be presumably recovered as Kaluza-Klein
compactification of Yang-Mills theory on an appropriate discrete
internal space.
 \sk
 ii) field theories on discrete structures are
interesting {\sl per se}: many statistical models are of this sort
and the tools offered by differential calculi on these structures
can be of use in the study of integrable models, see for ex. ref.
 \cite{DMintegrable}.
 \sk
 iii) Gauge
and gravity theories on finite group spaces may be used as lattice
approximations. For example the action for pure Yang-Mills $\int F
\we {}^* F$ considered on the finite group space $Z^N \times Z^N
\times Z^N \times Z^N$, yields the usual Wilson action of lattice
gauge theories, and $N \rightarrow \infty$ gives the continuum
limit \cite{DMgauge}. New lattice theories can be found by choosing
different finite groups.
\sk
Here we  propose an action for a toy theory of
gravity on the smallest nonabelian finite group $S_3$. In fact the
same type of action can be used for any finite group. Taking $Z^N
\times Z^N \times Z^N \times Z^N$  yields a discretized version of
gravity, in the same spirit of ref.s \cite{DMgrav} ( where however
 no action principle was used).
 \sk
   In Section 2 a review of the differential calculus
on finite groups is presented. Most of this material is not new,
and draws on the treatment of ref.s \cite{DMGcalculus,FI1}, where the Hopf
algebraic approach of Woronowicz \cite{Wor} for the construction of
differential calculi is adapted to the setting of finite groups.
Some developments on Lie derivative, diffeomorphisms and
integration are new. The general theory is illustrated in the case
of $S_3$ in Section 3. The ``softening" of the rigid finite group
manifold is discussed in Sect. 4, together with the application to
a gravity-like field theory on $S_3$.

%%%%%%%%%%%%%%%%%%%%%%%%%%%%%%
\sect{Differential calculus on finite groups}

\subsection{Fun(G) as a Hopf algebra}

Let $G$ be a finite group of order $n$ with generic element
$g$ and unit $e$. Consider $Fun(G)$,
the set of complex functions on $G$. An element $f$ of
$Fun(G)$ is specified by its values $f_g \equiv f(g)$
on the group elements $g$, and can be written as
\eq
f=\sum_{g \in G} f_g x^g,~~~f_g \in \Cb
\en
where the functions $x^g$ are defined by
\eq
x^g(g') = \de^g_{g'}
\en
Thus $Fun(G)$ is a n-dimensional vector space, and the $n$ functions
$x^g$ provide a basis. $Fun(G)$ is also a commutative algebra,
with the usual pointwise sum and product [$(f+h)(g)=f(g)+h(g)$,
$(f\cdot h)(g)=f(g)h(g)$, $ (\la f)(g)=\la f(g), f,h \in Fun(G),
\la \in \Cb$] and unit $I$ defined
by $I(g)=1, \forall g \in G$. In particular:
\eq
x^g x^{g'}=\de_{g,g'} x^g,~~~\sumong x^g = I \label{mul}
\en
The $G$ group structure induces a Hopf algebra structure on
$Fun(G)$, with  coproduct $\D$,  coinverse $\kappa$ and counit
$\epsi$ defined by group multiplication, inverse and unit as: \eqa
& &\D (f) (g,g') = f(gg'),~~~\D:Fun(G) \rightarrow Fun(G) \otimes
Fun(G)\\ & &\kappa (f)(g) = f(g^{-1}),~~~~~~~\kappa: Fun(G)
\rightarrow Fun(G) \\ & &\epsi (f)=f(e),~~~~~~~~~~~~~~\epsi:
Fun(G) \rightarrow \Cb \ena In the first line we have used $Fun(G
\times G) \approx Fun(G) \otimes Fun(G)$ [indeed a basis for
functions on $G \times G$ is given by $x^{g_1} \otimes x^{g_2},
g_1,g_2 \in G$]. On the basis functions $x^g$ the costructures
take the form: \eq \D (x^g)=\sum_{h \in G} x^h \otimes
x^{h^{-1}g},~~\kappa (x^g)=x^{g^{-1}},~~\epsi (x^g)=\de^g_e
\label{cox}
 \en
 The coproduct is related to the pullback induced
by left or right multiplication of $G$ on itself. Consider the
left multiplication by $g_1$: \eq L_{g_1}g_2=g_1g_2,~~~\forall
g_1,g_2 \in G
\en
This induces the left action (pullback) $\Lcal_{g_1}$ on
$Fun(G)$:
\eq
\Lcal_{g_1} f(g_2) \equiv f(g_1g_2)|_{g_2},~~~\Lcal_{g_1}:Fun(G)
\rightarrow Fun(G)
\en
where $f(g_1g_2)|_{g_2}$ means $f(g_1g_2)$ seen as a function
of $g_2$. For the basis functions we find easily:
\eq
\Lcal_{g_1} x^{g} = x^{g_1^{-1} g}
\en
Introducing the mapping
$\Lcal:Fun(G) \rightarrow Fun(G \times G)\approx Fun(G)
\otimes Fun(G)$:
\eq
(\Lcal f)(g_1,g_2) \equiv (\Lcal_{g_1}f)(g_2)=f(g_1g_2)|_{g_2}
\en
we see that
\eq
\Lcal = \D
\en
Thus the coproduct mapping $\D$ on the function $f$ encodes the
information on all the left actions $\Lcal_g, g \in G$ applied to
$f$, without reference to a particular $g$ (``point of the group
manifold"). It also encodes the information on right actions.

Indeed one can define the right action $\Rcal$ on $Fun(G)$ as:
 \eq
(\Rcal f)(g_1,g_2) \equiv (\Rcal_{g_1}f)(g_2)= f(g_2g_1)|_{g_2}
\en
 Introducing the flip operator $\tau : Fun(G\times G)
\rightarrow Fun(G \times G)$:
\eq
(\tau u)(g_1,g_2) \equiv u(g_2,g_1),~~~u \in Fun(G\times G)
\en
it is easy to find that:
\eq
\Rcal = \tau \circ \D
\en
For the basis functions:
\eq
\Rcal_{g_1} x^{g} = x^{g g_1^{-1}},~~\Rcal x^g=
\tau \circ\D (x^g)= \sum_{h \in G} x^{h^{-1}g} \otimes
x^h
\en
Finally:
\eq
\Lcal_{g_1} \Lcal_{g_2}=\Lcal_{g_1g_2},~~\Rcal_{g_1} \Rcal_{g_2}=\Rcal_{g_2g_1},~~
\Lcal_{g_1} \Rcal_{g_2}=\Rcal_{g_2} \Lcal_{g_1}
\en

\subsection{First order differential calculus}

Differential calculi can be constructed on Hopf algebras $A$ by
algebraic means, using the costructures of $A$ \cite{Wor}.
In the case of finite groups $G$, differential calculi on $A=Fun(G)$ have been
discussed in ref.s \cite{DMGcalculus,FI1}. Here we review some of the results,
and present new developments.
\sk
A {\bf {\sl first-order differential calculus}} on $A$ is defined by
\sk
i) a linear map $d$: $A \rightarrow \Gamma$, satisfying the Leibniz rule
\eq
d(ab)=(da)b+a(db),~~\forall a,b\in A; \label{Leibniz}
\en
The ``space of 1-forms" $\Ga$ is an
appropriate bimodule on $A$, which
essentially means that its elements can be
multiplied on the left and on the right by elements of $A$
[more precisely $A$ is a left module if $\forall a,b \in A, \forall
\rho,\rho' \in \Ga $ we have: $ a(\rho+\rho')=a\rho+a\rho',
~(a+b)\rho=a\rho+b\rho, ~a(b\rho)=(ab)\rho,~ I\rho=\rho$. Similarly
one defines a right module. A left and right module
is a {\sl bimodule} if
$a(\rho b)=(a\rho)b$]. From the Leibniz rule
$da=d(Ia)=(dI)a+Ida$ we deduce $dI=0$.
\sk
ii) the possibility of expressing any $\rho \in \Ga$ as
\eq
\rho=\sum_k a_k db_k \label{adb}
\en
\noi for some $a_k,b_k$ belonging to $A$. \sk To build a first
order differential calculus on $Fun(G)$ we need to extend the
algebra $A=Fun(G)$ to a differential algebra of elements
$x^g,dx^g$ (it is sufficient to consider the basis elements and
their differentials). Note however that the $dx^g$ are not
linearly independent. In fact from $0=dI=d(\sumong x^g)=\sumong
dx^g$ we see that only $n-1$ differentials are independent. Every
element $\rho = adb$ of $\Ga$ can be expressed as a linear
combination (with complex coefficients) of terms of the type $x^g
dx^{g'}$. Moreover $\rho b \in \Ga$ (i.e. $\Ga$ is also a right
module) since the Leibniz rule and the multiplication rule
(\ref{mul}) yield the commutations: \eq dx^g x^{g'} = -x^g
dx^{g'}+\de^g_{g'} dx^g
\en
allowing to reorder functions to the left of differentials.
There are $n(n-1)$ independent terms $x^g dx^{g'}$, since
there are $n-1$ independent $dx^g$. A convenient independent set
was chosen in ref. \cite{DMGcalculus} by taking all the terms
$e^{g,g'} \equiv x^g dx^{g'}$ with
$g \not= g'$. Within this set one can choose any subset, defining
a consistent first order differential algebra of elements
$x^g, e^{g,g'}$. These different choices can be described
by oriented graphs, whose vertices are elements $g$ of $G$,
and where an oriented line from $g$ to $g'$ means that
the term $x^g dx^{g'}$ exists in the subset \cite{DMGcalculus}.
\sk
{\bf Partial derivatives, ``curved" indices}
\sk
Consider the differential of a function $f \in Fun(g)$:
\eq
df = \sumong f_g dx^g = \sumongnote f_g dx^g + f_e dx^e=
\sumongnote (f_g - f_e)dx^g \equiv \sumongnote
\part_g f dx^g \label{partcurved}
\en
We have used $dx^e = - \sumongnote dx^g$ (from $\sumong dx^g=0$).
The partial derivatives of $f$ have been defined in
analogy with the usual differential calculus, and are given by
\eq
\part_g f = f_g - f_e = f(g) - f(e) \label{partcurved2}
\en
Not unexpectedly, they take here the form of finite differences
(discrete partial derivatives at the origin $e$). The partial
derivatives satisfy the modified Leibniz rule:
\eq
\part_g (ff')=(\part_g f) f'(g)-f(e) \part_g f'
\en

\subsection{ Left and right covariance}

 A differential calculus is
 left or right covariant if the left or right action of
 $G$ ($\Lcal_g$ or $\Rcal_g$) commutes with the exterior derivative $d$.
 Requiring left and right covariance in fact {\sl defines} the action of
 $\Lcal_g$ and $\Rcal_g$ on differentials: $\Lcal_g db \equiv
 d(\Lcal_g b), \forall b \in Fun(G)$ and similarly for
 $\Rcal_g db$. More generally, on elements of $\Ga$
 (one-forms) we define $\Lcal_g$ as:
 \eq
 \Lcal_g (adb) \equiv (\Lcal_g a) \Lcal_g db =
 (\Lcal_g a) d (\Lcal_g b)
 \en
 and similar for $\Rcal_g$.
 For example the left and right action on the differentials
 $dx^g$ is given by:
 \eq
\Lcal_g (dx^{g_1})\equiv d(\Lcal_g x^{g_1})=dx^{g^{-1}g_1},~~
\Rcal_g (dx^{g_1})\equiv d(\Rcal_g x^{g_1})=dx^{g_1 g^{-1}}
\en
In the same spirit as in the previous section, we can
 introduce mappings  $\DL:\Ga \rightarrow
A \otimes \Ga$ and $\DR: \Ga \rightarrow \Ga \otimes A$
that encode the information about all
left or right translations:
\eq
 \DL(a\rho b)=\D(a)\DL (\rho)\D(b),~~~\DL (db)=(id \otimes d) \D (b)
 ~~~ \forall a,b \in A,~\rho \in \Ga \label{leftco}
\en
\eq
 \DR(a\rho b)=\D(a)\DR (\rho)\D(b),~~~\DR (db)=(d \otimes id) \D (b)
 ~~~ \forall a,b \in A,~\rho \in \Ga \label{rightco}
\en
To see their relation with $\Lcal_g$ and $\Rcal_g$, consider
their action on the basic terms
$x^{g_1}dx^{g_2} \in \Ga$:
\eq
\DL (x^{g_1}dx^{g_2})=\D(x^{g_1})(id \otimes d) \D (x^{g_2})=
\sumonh x^{h}\otimes x^{h^{-1}g_1} dx^{h^{-1}g_2}\label{DLxdx}
\en
\eq
\DR (x^{g_1}dx^{g_2})=\D(x^{g_1})(d \otimes id) \D (x^{g_2})=
\sumonh x^{g_1 h} dx^{g_2 h} \otimes x^{h^{-1}} \label{DRxdx}
\en
Defining $(f \otimes \rho )[g] \equiv f(g) \rho \equiv
(\rho \otimes f )[g]$, with
$f \in A=Fun(G), \rho \in \Ga, g \in G$, we deduce:
\eq
\DL (x^{g_1}dx^{g_2})[g] = x^{g^{-1}g_1} dx^{g^{-1}g_2}=\Lcal_g
(x^{g_1}dx^{g_2}) \label{Lgxdx}
\en
\eq
\DR (x^{g_1}dx^{g_2})[g] = x^{g_1 g^{-1}} dx^{g_2 g^{-1}}=\Rcal_g
(x^{g_1}dx^{g_2}) \label{Rgxdx}
\en
so that the relations we were looking for are simply
\eq
\DL (\rho) [g] = \Lcal_g \rho,~ \DR (\rho) [g] = \Rcal_g \rho,
 ~ \forall \rho \in \Ga. \label{DLLcal}
\en
Computing $\DL$ and $\DR$ on the basic
differentials yields:
\eqa
& & \DL (dx^{g_1}) \equiv (id \otimes d) (\D x^{g_1})=
\sum_{h \in G} x^h \otimes dx^{h^{-1}g_1}\\
& & \DR (dx^{g_1}) \equiv (d \otimes id) (\D x^{g_1})=
\sum_{h \in G} dx^h \otimes x^{h^{-1}g_1}
\ena
In the following we will mainly use the pullbacks $\Lcal_g$ and
$\Rcal_g$, rather than the more cumbersome mappings $\DL$ and $\DR$.
The reason we have introduced them is to make contact with the
notations of general Hopf algebras, where the notion of ``point on
the manifold" may not exist. The reader not interested in Hopf
algebra formalism can simply ignore the discussions involving $\DL$
or $\DR$.
\sk
A differential calculus is called {\sl bicovariant} if it is both left and
right covariant.
\sk
Finally, consider the left action on $e^{g_1,g_2}=x^{g_1}dx^{g_2}$
given by eq. (\ref{Lgxdx}). We see that excluding from the
differential algebra the element $e^{g_1,g_2}$ implies the exclusion
of all the elements $e^{hg_1,hg_2}, h \in G$, that is the exclusion
of a subset of $xdx$ elements corresponding to an orbit under
left multiplication of the couples $(g_1,g_2)$. We call this subset
the $e^{g_1,g_2}$ left orbit.
 Thus the left-covariant
differential calculi on $Fun(G)$ are obtained from the
universal one (where none of the $e^{g_1,g_2}$ is excluded)
by excluding one or more  $e^{g_1,g_2}$ left orbits \cite{DMGcalculus}.
Analogous considerations hold for right-invariant calculi.

\subsection{Left and right-invariant one-forms}

As for Lie group manifolds, also in the case of finite groups one
can construct left and right invariant one-forms, which
 provide a basis (``vielbein basis" or cotangent basis)
for the vector space $\Ga$ of one-forms.
Following the usual definition, {\sl left-invariant}
one forms $\theta$ are elements of $\Ga$ satisfying:
\eq
\Lcal_g \theta = \theta
\en
In terms of the $\DL$ mapping this means:
\eq
\DL \theta = I \otimes \theta
\en
(use (\ref{DLLcal} and $I(g)=1$). It is a simple matter, via eq.
(\ref{DLxdx}), to show that
the  one-forms:
\eq
\theta^g \equiv \sumonh x^{hg} dx^h ~~(=\sumonh x^h dx^{hg^{-1}}),
\label{deftheta}
\en
are indeed left-invariant: $\DL \theta^g = I \otimes \theta^g$,
or equivalently : $\Lcal_h \theta^g=\theta^g$.

The relations (\ref{deftheta}) can be inverted :
 \eq
  dx^h = \sumong (x^{hg} - x^h)\theta^g \label{dxastheta}
\en
{} From $\sumong dx^g=0$ one finds:
\eq
\sumong \theta^g = \sumong
\sumonh x^h dx^{hg^{-1}}= \sumonh x^h \sumong dx^{hg^{-1}}=0
\label{sumtheta}
\en
We can take as basis of the cotangent space $\Ga$ the $n-1$
linearly independent left-invariant one-forms $\theta^g$
with $g \not= e$. This basis corresponds to the
``universal" differential calculus \cite{DMGcalculus}.
Smaller sets of $\theta^g$ can be chosen as basis (see below).

Notice that in the definition of $\theta^g$ the whole
$e^{g,e}$ orbit is involved, cf. (\ref{deftheta}).
Thus the left-invariant one-forms are in 1-1 correspondence
with the (g,e) left orbits: removing the $e^{g,e}$ left orbit
means to remove $\theta^g$. All left-covariant differential
calculi are therefore obtained by excluding (i.e. setting to
zero) some of the $\theta^g$.

The remaining $\theta^g$ ($g \not= e$) constitute a basis for the bimodule
$\Ga$. The $x,~\theta$ commutations (bimodule relations)
are easily found:
\eq
x^h dx^g = x^h \theta^{g^{-1}h} =
\theta^{g^{-1}h} x^g ~~(h\not=g)~~\Rightarrow x^h \theta^g =
\theta^g x^{hg^{-1}}~~(g\not=e)
\label{xthetacomm}
\en
implying the general commutation rule between functions
and left-invariant one-forms:
\eq
f \theta^g = \theta^g \Rcal_g f \label{fthetacomm}
\en
Thus functions do commute between themselves (i.e. $Fun(G)$ is
a commutative algebra) but do not commute with the
basis of one-forms $\theta^g$. In this sense the differential geometry
of $Fun(G)$ is noncommutative, the noncommutativity being
milder than in the case of quantum groups $Fun_q(G)$(which
are noncommutative algebras).
\sk

Analogous results hold for right invariant one-forms $\om^g$,
the corresponding formulae being:
\eq
\om^g = \sumonh x^{gh}dx^h,~~\DR\om^g = \om^g \otimes I
\en
\eq
f\om^g = \om^g \Lcal_g f
\en
From the expressions of $\theta^g$ and $\om^g$
in terms of $xdx$, one finds the relations
\eq
\theta^g = \sumonh x^h \om^{ad(h)g},~~~\om^g=\sumonh x^h
\theta^{ad(h^{-1})g}
\en
For a bicovariant calculus the right action on $\theta^g$
is given by (use the definitions of $\DR$ and of $\theta^g$):
\eq
\DR \theta^g= \sumonh \theta^{ad(h) g}\otimes x^h,~~\mbox{or}~~
\Rcal_h  \theta^g = \theta^{ad(h) g}
\en
where $ad$ is the adjoint action of $G$ on $G$, i.e. $ad(h)g
\equiv hgh^{-1}$. Then {\sl bicovariant calculi are in 1-1
correspondence with unions of conjugacy classes (different from
$\{e\}$)} \cite{DMGcalculus}: if $\theta^g$ is set to zero, one must set to
zero all the $\theta^{ad(h)g},~\forall h \in G$ corresponding
to the whole conjugation class of $g$.
 \sk
 As in \cite{DMGcalculus} we denote by $G'$ the subset corresponding
 to  the union of conjugacy classes
 that characterizes the bicovariant calculus on $G$
 ($G' = \{g \in G |\theta^g \not= 0\}$).
 Unless otherwise indicated, hereafter repeated indices are
 summed on $G'$.
 \sk

A bi-invariant (i.e. left and right invariant)
one-form $\Theta$ is obtained by summing
on all $\theta^g$ or $\om^g$ with $g \not= e$:
\eq
\Theta = \sumongnote \theta^g = \sumongnote \om^g
\en

{\bf Note 2.4.1}: since $\sumong \theta^g=0=\sumong \om^g$,
 cf. (\ref{sumtheta}), we have
also $\theta^e = -\Theta= \om^e$.

\subsection{Exterior product}

For a bicovariant differential calculus on a Hopf
algebra $A$ an {\sl exterior product}, compatible with the left
and right actions of $G$, can be defined by means of a bimodule
automorphism
$\La$ in $\Gamma \otimes \Gamma$ that generalizes the
ordinary permutation operator:
\eq
\La (\theta \otimes \om )= \om \otimes \theta, \label{La}
\en
\noi where $\theta$ and $\om$ are respectively left and right
invariant elements of $\Ga$ \cite{Wor}. Bimodule automorphism means that
\eq
\La(a\eta)=a\La (\eta) \label{bimauto1}
\en
\eq
\La(\eta b)=\La(\eta)b \label{bimauto2}
\en
\noi for any $\eta \in \Ga \otimes \Ga$ and $a,b \in A$.
The tensor product between elements $\rho,\rhop \in \Ga$
is defined to
have the properties $\rho a\otimes \rhop=\rho \otimes a \rhop$, $a(\rho
\otimes \rhop)=(a\rho) \otimes \rhop$ and $(\rho \otimes \rhop)a=\rho
\otimes (\rhop a)$.

 Left and right actions on $\Ga \otimes \Ga$ are
defined by
 \footnote{ More generally, we can define the action of $\DL$
on $\Ga \otimes \Ga \otimes \cdots \otimes \Ga$ as

\[
\DL (\rho \otimes \rhop \otimes \cdots \otimes \rhopp)\equiv
\rho_1 \rhop_1 \cdots \rhopp_1 \otimes \rho_2 \otimes
\rhop_2\otimes \cdots \otimes \rhopp_2
\]
\[
\DL: \Ga \otimes \Ga \otimes \cdots\otimes \Ga \rightarrow
A\otimes\Ga\otimes\Ga\otimes\cdots\otimes\Ga
\]
\[
\DR (\rho \otimes \rhop \otimes \cdots \otimes \rhopp)\equiv
\rho_1 \otimes \rhop_1 \cdots \otimes \rhopp_1 \otimes \rho_2
\rhop_2 \cdots \rhopp_2
\]
\[
\DR: \Ga \otimes \Ga \otimes \cdots\otimes \Ga \rightarrow
\Ga\otimes\Ga\otimes\cdots\otimes\Ga\otimes A.
\]
}:
\eq
\DL (\rho \otimes \rhop)\equiv \rho_1 \rhop_1 \otimes \rho_2 \otimes
\rhop_2,~~~\DL: \Ga \otimes \Ga \rightarrow A\otimes\Ga\otimes\Ga
\label{DLGaGa}
\en
\eq
\DR (\rho \otimes \rhop)\equiv \rho_1 \otimes \rhop_1 \otimes \rho_2
\rhop_2,~~~\DR: \Ga \otimes \Ga \rightarrow \Ga\otimes\Ga\otimes A
\label{DRGaGa}
\en
\noi where $\rho_1$, $\rho_2$, etc., are a customary short-hand notation
defined by
\eq
\DL (\rho) = \rho_1 \otimes \rho_2,~~~\rho_1\in A,~\rho_2\in \Ga
\en
\eq
\DR (\rho) = \rho_1 \otimes \rho_2,~~~\rho_1\in \Ga,~\rho_2\in A.
\en

\noi Left-invariance on $\Ga\otimes\Ga$ is naturally defined as
$\DL (\rho \otimes \rhop) = I \otimes \rho \otimes \rhop$ (similar
definition for right-invariance), so that, for example, $\theta^i
\otimes \theta^{j}$ is left-invariant, and is in fact a
left-invariant basis for $\Ga \otimes \Ga$ if $\{ \theta^i \}$ is
a left-invariant basis for $\Ga$.

The definition of $\Lcal_g$ and
$\Rcal_g$ on tensor products $\Ga \otimes ... \otimes \Ga$ is
straightforward; for example:
 \eq
 \Lcal_g (\rho \otimes \rhop)\equiv \DL (\rho \otimes \rhop)[g]
 =\rho_1 \rhop_1 (g) \rho_2 \otimes \rhop_2=\Lcal_g \rho \otimes
 \Lcal_g \rhop
 \en
\eq
 \Rcal_g (\rho \otimes \rhop)\equiv \DR (\rho \otimes \rhop)[g]
 =\rho_1 \otimes  \rhop_1  \rho_2 \rhop_2 (g)=
 \Rcal_g \rho \otimes
 \Rcal_g \rhop
 \en
 where the last equality in both equations is derived
 after expanding the generic form $\rho$ on the $\theta^i$
 basis ($\rho=f_i \theta^i$) and likewise for $\rhop$.
 In particular $\Lcal_h (\theta^i \otimes \theta^j)=
\theta^i \otimes \theta^j$, $\Rcal_h (\theta^i \otimes \theta^j)=
\theta^{ad(h)i} \otimes \theta^{ad(h)j}$.
 \sk
 {\bf Note 2.5.1}: In general $\La^2 \not= 1$, since $\La(\om^{j} \otimes
\theta^i )$ is not necessarily equal to $ \theta^i \otimes \om^{j}
$. By linearity, $\La$ can be extended to the whole of $\Gamma
\otimes \Gamma$.
\sk
{\bf Note 2.5.2}:
$\La$ is invertible and commutes with the left and right action of
$G$, i.e. $\DL \La (\rho \otimes \rhop)=(id \otimes \La) \DL
(\rho \otimes \rhop)
= \rho_1\rhop_1 \otimes \La(\rho_2 \otimes \rhop_2)$, and
similar for $\DR$. Equivalently: $\Lcal_g \La (\rho \otimes \rhop)=
\La [\Lcal_g(\rho \otimes \rhop)]=
 \La (\Lcal_g\rho \otimes \Lcal_g \rhop)$ and similar for $\Rcal_g$.
 Therefore $\La
(\theta^{i} \otimes \theta^{j})$ is
left-invariant, and can be expanded on the left-invariant
basis $\theta^{i} \otimes \theta^{j}$:
\eq
\La (\theta^{i} \otimes \theta^{j})= \L{ij}{kl}
\theta^{k} \otimes \theta^{l}.
\en
\sk
The {\sl exterior product}
is defined as:
\eqa
& & \rho \we \rho ' \equiv \rho \otimes \rho ' -
\La (\rho \otimes \rho ')\\
& & \theta^i \we \theta^j \equiv \W{ij}{kl} \theta^k \otimes \theta^l=
 \theta^i \otimes \theta^j -
\L{ij}{kl} \theta^k \otimes \theta^l. \label{extheta}
\ena
\noi where $\rho,\rhop \in \Ga$ and $\{ \theta^i \}$ = left-invariant basis for
$\Ga$.  Notice that, given the matrix $\L{ij}{kl}$, we can compute the
exterior product of any $\rho,\rhop \in \Ga$, since
any $\rho \in \Ga$ is
expressible in terms of $\theta^i$.
\sk

In the case $A=Fun(G)$, we find
 \eqa
  & &\La (\theta^g \otimes
\theta^{g'})=\La (\theta^g \otimes \sumonh x^h \om^{ad(h)g'}) =
\sumonh x^{hg} \La (\theta^g \otimes \om^{ad(h)g'})= \nonumber \\
& &~~~~~~~~~~~~~~= ~\sumonh x^{hg} \om^{ad(h)g'} \otimes \theta^g
= \sum_{h,h'\in G} x^{hg} x^{h'} \theta^{ad(h'^{-1}h)g'}\otimes
\theta^g= \nonumber \\ & & ~~~~~~~~~~~~~~ = \sumonh x^{hg}
\theta^{ad(g^{-1})g'} \otimes \theta^g =\theta^{ad(g^{-1})g'}
\otimes \theta^g
 \ena
  and the $\L{ij}{kl}$ matrix takes the form:
\eq
 \L{g_1,g_2}{h_1,h_2} = \de^{g_1}_{h_2}
\de^{ad(g_1^{-1})g_2}_{h_1}
\en
Then the exterior product of two left-invariant basic one-forms is
given by:
\eq
\theta^{g_1} \we \theta^{g_2}=\theta^{g_1} \otimes \theta^{g_2}
- \theta^{g_1^{-1} g_2 g_1} \otimes \theta^{g_1}
\en
Note that:
 \eq
  \theta^{g} \we \theta^{g}=0~~~~\mbox{(no sum on $g$)}
\en
 This familiar formula holds for $Fun(G)$, but not
for a general Hopf algebra.
 \sk
 We can generalize the definition
(\ref{extheta}) to exterior products of $n$ one-forms:
 \eq
\theta^{i_1} \we ... \we \theta^{i_n} \equiv \W{i_1..i_n}{j_1..j_n}~
 \theta^{j_1} \otimes ...\otimes
\theta^{j_n}
\en
\noi or in short-hand notation:
\eq
\theta^{1} \we ... \we \theta^{n}= W_{1..n}~
 \theta^{1} \otimes ...\otimes
\theta^{n}
\en
\noi where the labels 1..n in $W$ refer to index couples.
The numerical coefficients $W_{1\ldots n}$ are given through a recursion
relation
\eq
W_{1\ldots n} = {\cal I}_{1\ldots n} W_{1\ldots n-1} , \label{wedge2}
\en
where
\eq
{\cal I}_{1\ldots n} = 1 - \Lambda_{n-1,n} +
\Lambda_{n-2,n-1} \Lambda_{n-1,n}   \ldots
-(-1)^n \Lambda_{12} \Lambda_{23} \cdots \Lambda_{n-1,n}
\label{wedge3}
\en
and $W_1 = 1$. The space of $n$-forms $\Ga^{\we n}$ is therefore defined as in
the usual case but with the new permutation operator $\La$,
and can be shown to be a bicovariant bimodule
(see for ex. \cite{Athesis}), with left
and right action defined as for $\Ga \otimes ...\otimes \Ga$
with the tensor product replaced by the wedge product.

\subsection{Exterior derivative}

 With the exterior product we can define the {\sl exterior
derivative}
\eq
d~:~\Gamma \rightarrow \Gamma \we \Gamma
\en
\eq
d (a_k db_k) = da_k \we db_k,
\en
\noi which can easily be extended to $\Gamma^{\we n}$ ($d:
\Gamma^{\we n} \rightarrow \Gamma^{\we (n+1)}$), and has the
following properties:
\eq
 d(\rho \we \rhop)=d\rho \we \rhop +
(-1)^k \rho \we d\rhop \label{propd1}
\en
\eq
d(d\rho)=0\label{propd2}
\en
\eq
\DL (d\rho)=(id\otimes d)\DL(\rho)~~
\mbox{or}~~\Lcal_g (d\rho)=d ( \Lcal_g \rho) \label{propd3}
\en
\eq
\DR (d\rho)=(d\otimes id)\DR(\rho)~~
\mbox{or}~~\Rcal_g (d\rho)=d ( \Rcal_g \rho),
\label{propd4}
\en
\noi where $\rho \in \Ga^{\we k}$, $\rhop \in \Ga^{\we n}$,
$\Ga^{\we 0} \equiv Fun(G)$. The
last two properties show that $d$ commutes with the
left and right action of $G$.
\sk

\subsection{Tangent vectors}

In  (\ref{partcurved}) we expressed $df$ in terms of the
differentials $dx^g$. Using (\ref{dxastheta}) we find the
expansion of $df$ on the basis of the left-invariant one-forms
$\theta^g$:
 \eq
  df=\sumong f_g dx^g = \sumong f_g \sumonhp
(x^{gh}-x^g) \theta^h = \sumonhp (\Rcal_{h^{-1}} f - f ) \theta^h
\equiv \sumonhp (t_h f) \theta^h \label{partflat}
\en
so that the ``flat" partial derivatives $t_h f$ are given by \eq
t_h f = \Rcal_{h^{-1}} f - f \label{partflat2}
\en
Note that $t_h f$ are really functions $\in Fun(G)$, whereas the
``curved" partial derivatives of eq. (\ref{partcurved2}) are
numbers. The Leibniz rule for the flat partial derivatives $t_g$
reads:
\eq
 t_g (ff')=(t_g f) \Rcal_{g^{-1}} f'  + f t_g f' \label{tg}
\en

In analogy with ordinary differential calculus, the operators
$t_g$ appearing in (\ref{partflat}) are called (left-invariant)
{\sl tangent vectors}, and in our case are given by
 \eq
  t_g =
\Rcal_{g^{-1}}- id \label{tangent}
\en
They satisfy the composition rule:
 \eq
  t_g t_{g'}= \sum_h
\c{h}{g,g'} t_h \label{chichi}
\en
where the structure constants are:
\eq
\c{h}{g,g'}=\de^h_{g'g} - \de^h_{g}-\de^h_{g'}
\label{cconst}
\en
and have the property:
\eq
\c{ad(h)g_1}{~~ad(h)g_2,ad(h)g_3}= \c{g_1}{g_2,g_3} \label{adhc}
\en

Clearly we can expand $df$ also on the right-invariant basis
$\om^g$ and define (right-invariant) tangent vectors $\tti_h$ from
$df=\sum_h (\tti_h f) \om^h$, whose explicit operator expression
is:
 \eq
 \tti_g = \Lcal_{g^{-1}}- id
\en

{\bf Note 2.7.1 :}
The exterior derivative on any $f \in Fun(G)$ can be expressed as
a commutator of $f$ with the bi-invariant one-form $\Theta$:
\eq
df = [\Theta , f]
\en
as one proves by using (\ref{fthetacomm}) and (\ref{partflat}).
\sk
 {\bf Note 2.7.2 :} From the fusion rules (\ref{chichi}) we
deduce the ``deformed Lie algebra"
(cf. ref.s  \cite{Wor,ACintro,LCqISO,Athesis}):
  \eq
  t_{g_1} t_{g_2} -
\L{g_3,g_4}{g_1,g_2}t_{g_3} t_{g_4}= \C{h}{g_1,g_2} t_h
\en
where the $\Cb$ structure constants are given by:
 \eq
\C{g}{g_1,g_2} \equiv \c{g}{g_1,g_2} - \L{g_3,g_4}{g_1,g_2}
\c{g}{g_3,g_4}= \c{g}{g_1,g_2} -
\c{g}{g_2,g_2 g_1 g_2^{-1}}=
\de^{ad(g_2^{-1})g}_{g_1} - \de^g_{g_1}
\label{Cconst}
\en
and besides property (\ref{adhc}) they also satisfy:
 \eq
\C{g}{g_1,g_2}=\C{g_1}{g,g_2^{-1}} \label{propC}
\en
Moreover the following identities hold:

{\bf i)} {\sl deformed Jacobi identities:}
\eq \C{k}{h_1,g_1}
\C{h_2}{k,g_2} - \L{g_3,g_4}{g_1,g_2}\C{k}{h_1,g_3}\C{h_2}{k,g_4}=
\C{k}{g_1,g_2} \C{h_2}{h_1,k} \label{Jacobi}
\en

{\bf ii)} {\sl fusion identities:} \eq
\C{k}{h_1,g} \C{h_2}{k,g'}=
\c{h}{g,g'} \C{h_2}{h_1,h} \label{adfusion}
\en

Thus the $\Cb$ structure constants are a representation (the
adjoint representation) of the tangent vectors $t$.
 \sk
 {\bf Note 2.7.3:} The fusion rules (\ref{chichi}) also allow to associate
 an ordinary
(i.e. not deformed) Lie algebra to the finite group $G$;
 the corresponding structure constants are simply
twice the antisymmetric part (in the indices $g_1,g_2$) of
$\c{g}{g_1,g_2}$.

\subsection{Cartan-Maurer equations, connection and curvature}

From the definition (\ref{deftheta}) and eq. (\ref{xthetacomm})
we deduce the Cartan-Maurer equations:
\eq
d\theta^g + \sum_{g_1,g_2} \c{g}{g_1,g_2}\theta^{g_1}\we
\theta^{g_2}=0 \label{CM}
\en
where the structure constants $\c{g}{g_1,g_2}$ are those
given in (\ref{cconst}).
\sk
Parallel transport of the vielbein $\theta^g$
can be defined as in
ordinary Lie group manifolds:
\eq
\nabla \theta^g= - \omc{g}{g'} \otimes \theta^{g'}
 \label{parallel}
\en
where $\omc{g_1}{g_2}$ is the connection one-form:
\eq
\omc{g_1}{g_2}= \Gc{g_1}{g_3,g_2} \theta^{g_3}
\en
Thus parallel transport is a map from $\Ga$ to $\Ga \otimes \Ga$;
by definition it must satisfy:
\eq
\nabla (a \rho) = (da)\otimes \rho + a \nabla \rho,~~~\forall a \in
A,~\rho \in \Ga \label{parallel1}
\en
and it is a simple matter to verify that this relation is
satisfied with the usual parallel transport of Riemannian
manifolds. As for the exterior differential, $\nabla$ can be
extended to a map $\nabla : \Ga^{\we n} \otimes \Ga
\longrightarrow \Ga^{\we (n+1)} \otimes \Ga $ by defining:
 \eq
\nabla (\varphi \otimes \rho)=d\varphi \otimes \rho +
 (-1)^n \varphi \nabla
\rho
\en

Requiring parallel transport to commute with the left and right
action of $G$ means:
 \eqa & &\Lcal_{h} (\nabla \theta^{g})=\nabla
( \Lcal_{h} \theta^{g}) =\nabla \theta^g\\ & &\Rcal_{h} (\nabla
\theta^{g})=\nabla ( \Rcal_{h} \theta^{g}) =\nabla \theta^{ad(h)g}
\ena
 Recalling that  $\Lcal_{h} (a \rho)=(\Lcal_h a) (\Lcal_h
\rho)$ and $\Lcal_{h} (\rho \otimes \rho')=(\Lcal_h \rho) \otimes
(\Lcal_h \rho'),~\forall a \in A,~\rho,~\rho' \in \Ga$ (and
similar for $\Rcal_h$),
 and substituting
(\ref{parallel}) yields respectively:
\eq
\Gc{g_1}{g_3,g_2} \in \Cb
\en
and
 \eq
 \Gc{ad(h)g_1}{ad(h)g_3,ad(h)g_2}=\Gc{g_1}{g_3,g_2} \label{adga}
\en
Therefore the same situation arises as in the case of Lie groups,
for which  parallel transport on the group manifold commutes
with left and right action iff the connection components are
$ad(G)$ - conserved constant tensors. As for Lie groups, condition
(\ref{adga}) is satisfied if one takes $\Ga$ proportional to the
structure constants. In our case, we can take any combination of
the $C$ or $\Cb$ structure constants, since both are $ad(G)$
conserved constant tensors. As we see below, the $C$ constants
can be used to define a torsionless connection, while the $\Cb$
constants define a parallelizing connection.

\sk
 As usual, the {\sl curvature} arises from $\nabla^2$:
  \eq
  \nabla^2 \theta^g = - \R{g}{g'} \otimes \theta^{g'}
\en
\eq
\R{g_1}{g_2} \equiv d \omc{g_1}{g_2} + \omc{g_1}{g_3} \we
\omc{g_3}{g_2} \label{curvature}
\en

The {\sl torsion} $R^g$ is defined by:
\eq
R^{g_1} \equiv d\theta^{g_1} +  \omc{g_1}{g_2} \we \theta^{g_2}
\label{torsion}
\en

Using the expression of $\om$ in terms of $\Ga$ and the
Cartan-Maurer equations yields
 \eqa
 \R{g_1}{g_2} &=& (-
\Gc{g_1}{h,g_2} \c{h}{g_3,g_4} + \Gc{g_1}{g_3,h} \Gc{h}{g_4,g_2})~
\theta^{g_3} \we \theta^{g_4}=\\
&=& (-
\Gc{g_1}{h,g_2} \C{h}{g_3,g_4} + \Gc{g_1}{g_3,h} \Gc{h}{g_4,g_2}-
\Gc{g_1}{g_4,h} \Gc{h}{g_4g_3g_4^{-1},g_2})~\theta^{g_3} \otimes
 \theta^{g_4}\nonumber
\ena
\eqa
R^{g_1}&=& (- \c{g_1}{g_2,g_3} + \Gc{g_1}{g_2,g_3})~ \theta^{g_2} \we
\theta^{g_3}= \nonumber \\
&=& (- \C{g_1}{g_2,g_3} + \Gc{g_1}{g_2,g_3}-
\Gc{g_1}{g_3,g_3g_2g_3^{-1}})~\theta^{g_2} \otimes
\theta^{g_3}
\ena

Thus a connection satisfying:
 \eq
  \Gc{g_1}{g_2,g_3}-
\Gc{g_1}{g_3,g_3g_2g_3^{-1}}=\C{g_1}{g_2,g_3} \label{rconn}
  \en
corresponds to a vanishing torsion $R^g =0$ and could be
  referred to as a ``Riemannian" connection.
\sk
 On the other hand,  the choice:
   \eq
  \Gc{g_1}{g_2,g_3}=\C{g_1}{g_3,g_2^{-1}} \label{parconn}
  \en
corresponds to a vanishing curvature $\R{g}{g'}=0$, as can be
checked by using the fusion equations (\ref{adfusion}) and
property (\ref{propC}). Then (\ref{parconn}) can be called the
parallelizing connection: {\sl finite groups are
parallelizable.}
\sk

\subsection{Tensor transformations and covariant derivative}

Under the familiar transformation of the connection 1-form:
\eq
(\omc{i}{j})' = \a{i}{k} \omc{k}{l} \ainv{l}{j} +
\a{i}{k} d \ainv{k}{j} \label{omtransf}
\en
the curvature 2-form transforms homogeneously:
\eq
(\R{i}{j})' = \a{i}{k} \R{k}{l} \ainv{l}{j}
\en
The transformation rule (\ref{omtransf}) can be seen as induced by
the change of basis $\theta^i=\a{i}{j} \theta^j$, with $\a{i}{j}$
invertible $x$-dependent matrix (use eq. (\ref{parallel1}) with
$a\rho=\a{i}{j} \theta^j$).
\sk
The covariant derivative $D$ of a
function
 $\phi_i$ transforming as $\phi_i'=\phi_j \ainv{j}{i}$
 ($i$ = contravariant index)  is defined as follows:
 \eq
 D \phi_i \equiv d\phi_i - \phi_j \omc{j}{i}
 \en
(or equivalently by $\nabla \phi \equiv D\phi_i \otimes \theta^i$,
with $\phi = \phi_i \theta^i$), and indeed transforms
homogeneously $(D \phi_i )' = (D \phi_j) \ainv{j}{i}$

Similarly on a function $\varphi^i$ with a covariant index
transforming as $(\varphi^i)'=\a{i}{j} \varphi^j$ the covariant
derivative is:
\eq
 D \varphi^i \equiv d\varphi^i + \omc{i}{j}
\varphi_j
\en
and transforms as $(D\varphi^i)'=\a{i}{j} (D\varphi^j)$.
 Then $D$
on the scalar $\phi_i \varphi^i$ reduces to $d$, if one defines
$D$ to satisfy the Leibniz rule: $D (\phi_i \varphi^i)=(D\phi_i )
\varphi^i+\phi_i D( \varphi^i)$.
 \sk
The generalization of $D$ on tensors $T$ with an arbitrary
  number of covariant and contravariant indices is not
  straightforward, see for example  ref. \cite{DMGcalculus,DMgrav} for a discussion.
  Although a consistent definition of parallel transport
  for tensors is clearly important, we will not need it in the following.
\sk

\subsection{Metric}

The metric tensor $\ga$ can be defined as an element of $\Ga \otimes
\Ga$:
 \eq
  \ga = \ga_{i,j} \theta^i \otimes \theta^j
  \en
 Requiring it to be invariant under left and right action of
 $G$ means:
 \eq
 \Lcal_h (\ga)=\ga=\Rcal_h (\ga)
 \en
or equivalently, by recalling $\Lcal_h(\theta^i \otimes
\theta^j)=\theta^i \otimes \theta^j$, $\Rcal_h(\theta^i \otimes
\theta^j)=\theta^{ad(h)i}\otimes \theta^{ad(h)j}$  :
 \eq
  \ga_{i,j} \in \Cb,~~  \ga_{ad(h)i,ad(h)j}=\ga_{i,j} \label{gabiinv}
 \en
These properties are  analogous to the ones satisfied by the
Killing metric of Lie groups, which is indeed constant and
invariant under the adjoint action of the Lie group.
 \sk
 On finite $G$ there are various choices of biinvariant
 metrics. One can simply take $\ga_{i,j}=\de_{i,j}$,
   or $\ga_{i,j}= \C{k}{l,i} \C{l}{k,j}$, or the
   ``distance" matrix defined in Section 3.
 Note that we are not insisting here on a covariantly conserved
metric (i.e. a metric compatible connection,
see ref. (\cite{DMGcalculus})).
\sk
 For any biinvariant metric $\ga_{i,j}$ there are tensor transformations
 $\a{i}{j}$ under which $\ga_{i,j}$ is invariant, i.e.:
 \eq
 \a{h}{h'} \ga_{h,k} \a{k}{k'}=\ga_{h',k'} \Leftrightarrow
\a{h}{h'} \ga_{h,k} = \ga_{h',k'} \ainv{k'}{k} \label{ginv}
 \en
 These transformations are simply given by the matrices that
rotate the indices according to the adjoint action of $G$:
 \eq
 \a{h}{h'} (g) = \de^{ad(\al(g))h}_{h'} \label{Gadjoint}
 \en
 where $\al(g): G \mapsto G$ is an arbitrary mapping. Then
 these matrices are functions of $G$ via this mapping, and
 their action leaves $\ga$ invariant because of its biinvariance
 (\ref{gabiinv}). Indeed
substituting these matrices in (\ref{ginv}) yields:
 \eq
 \a{h}{h'} (g) \ga_{h,k} \a{k}{k'} (g)=
 \ga_{ad([\al(g)]^{-1})h',ad([\al(g)]^{-1})k'}= \ga_{h',k'}
 \en
proving the invariance of $\ga$.

Consider now a contravariant vector $\varphi^i$ transforming as
$(\varphi^i)'=\a{i}{j}(\varphi^j)$. Then using (\ref{ginv}) one
can easily see that
 \eq
  (\varphi^k \ga_{k,i})'=  \varphi^{k'} \ga_{k',i'} \ainv{i'}{i}
  \en
  i.e.  the vector $\varphi_i \equiv \varphi^k \ga_{k,i}$ indeed
  transforms as a covariant vector.

\subsection{Lie derivative and diffeomorphisms}

The notion of diffeomorphisms, or general coordinate
transformations, is fundamental in gravity theories. Is there such
a notion in the setting of differential calculi on Hopf algebras ?
The answer is affirmative, and has been discussed in detail in
ref.s \cite{ACintro,LCqISO,Athesis}.
As for differentiable manifolds, it relies on the existence of the
Lie derivative.

Let us review the situation for Lie group manifolds. The Lie
derivative $l_{t_i}$ along a left-invariant tangent vector $t_i$
is related to the infinitesimal right translations generated by
$t_i$:
 \eq
 l_{t_i} \rho = \limepsizero {1\over \epsi} [\Rcal_{\exp [\epsi
 t_i]} \rho - \rho] \label{Liederivative1}
 \en
 $\rho$ being an arbitrary tensor field. Introducing the
 coordinate dependence
 \eq
 l_{t_i} \rho (y) = \limepsizero {1\over \epsi}
  [\rho (y + \epsi t_i) - \rho (y) ]
 \en
identifies the Lie derivative $ l_{t_i}$ as a directional
derivative along $t_i$.  Note the difference in meaning of the
symbol $t_i$ in the r.h.s. of these two equations: a group
generator in the first, and the corresponding tangent vector in
the second.

To find the natural generalization of the Lie derivative in the
case of finite groups, we express formula (\ref{Liederivative1})
in a completely algebraic notation:
 \eq
l_{t_i} \rho=(id \otimes t_i) \DR (\rho) \label{liederal}
\en
This expression is well defined for any Hopf algebra. In
particular for finite groups (\ref{liederal}) takes the form:
 \eq
 l_{t_g} \rho =  [\Rcal_{g^{-1}} \rho - \rho] \label{Liederivative2}
 \en
so that the Lie derivative is simply given by
 \eq
l_{t_g}=\Rcal_{g^{-1}}-id=t_g
\en
cf. the definition of $t_g$ in (\ref{tangent}). For example \
 \eq
l_{t_g} (\theta^{g_1} \otimes \theta^{g_2}) =
 \theta^{ad(g^{-1})g_1} \otimes \theta^{ad(g^{-1})g_2}-
 \theta^{g_1} \otimes \theta^{g_2}
 \en

 As in the case of differentiable manifolds, the Cartan formula
 for the Lie derivative acting on p-forms holds:
\eq
 l_{t_g}= i_{t_g} d + d i_{t_g}
 \en
(see Appendix A ).

 Exploiting this formula, diffeomorphisms
  (Lie derivatives) along generic tangent vectors $V$
 can also be consistently defined via the operator:
\eq
 l_{V}= i_{V} d + d i_{V}
 \en
 This requires
  a suitable definition
 of the contraction operator $i_V$  along generic tangent vectors
 $V$, discussed in Appendix A.

 We have then a way
 of defining ``diffeomorphisms" along arbitrary (and x-dependent)
 tangent vectors for any tensor $\rho$:
 \eq
 \delta \rho = l_V \rho
 \en
and of testing the invariance of candidate lagrangians under the
generalized Lie derivative.

\subsection{Haar measure and integration}

Since we want to be able to define actions (integrals on
$p$-forms) we must now define integration of $p$-forms on finite
groups.

 Let us start with integration of functions $f$. We define the integral
  map $h$ as a linear functional $h: Fun(G) \mapsto \Cb$ satisfying the
  left and right invariance conditions:
  \eq
  h(\Lcal_g f)=0=h(\Rcal_g f)
  \en
  Then this map is uniquely determined (up to a normalization constant),
  and is simply given by the ``sum over $G$" rule:
  \eq
  h(f)= \sumong f(g)
  \en

 Next we turn to define the integral of a p-form.
Within the differential calculus we have a basis of left-invariant
1-forms, which may allow the definition of a biinvariant volume
element. In general for a differential calculus with $n$
independent tangent vectors, there is an integer $p  \geq n$ such
that the linear space of $p$-forms is 1-dimensional, and $(p+1)$-
forms vanish identically. We will see explicit examples in the
next Section. This means that every product of $p$ basis
one-forms $\theta^{g_1} \we \theta^{g_2} \we ... \we \theta^{g_p}$
is proportional to one of these products, that can be chosen to
define the volume form $vol$:
 \eq
 \theta^{g_1} \we \theta^{g_2} \we ... \we \theta^{g_p}=
 \epsilon^{g_1,g_2,...g_p} vol
 \en
 where $\epsilon^{g_1,g_2,...g_p}$ is the proportionality constant.
 Note that the volume $p$-form is obviously left invariant. We can
  prove that it is also right invariant with the following
  argument. Suppose that $vol$ be given by
  $\theta^{h_1} \we \theta^{h_2} \we ... \we \theta^{h_p}$ where
  $h_1,h_2,...h_p$ are given group element labels. Then the right
  action on $vol$ yields:
  \eq
  \Rcal_g [\theta^{h_1} \we  ... \we
  \theta^{h_p}]=
  \theta^{ad(g)h_1} \we ... \we
  \theta^{ad(g)h_p}=
  \epsilon^{ad(g)h_1,...ad(g)h_p} vol
  \en
  Recall now that the ``epsilon tensor" $\epsilon$ is necessarily
  made out of products of the  $\La$ tensor of eq. (\ref{extheta}), defining the wedge
  product. This tensor is invariant under the adjoint action
  $ad(g)$, and so is the $\epsilon$ tensor. Therefore
  $\epsilon^{ad(g)h_1,...ad(g)h_p}=\epsilon^{h_1,...h_p}=1$
  and $\Rcal_g vol = vol$. This will be verified in the examples
  of next Section.

Having identified the volume $p$-form it is natural to set
 \eq
 \int f vol \equiv h(f) = \sumong f(g) \label{intpform}
 \en
 and  define the integral on a $p$-form $\rho$ as:
 \eq
  \int \rho = \int \rho_{g_1,...g_p}~ \theta^{g_1}
 \we ... \we \theta^{g_p}=   \int
\rho_{g_1,...g_p}~\epsilon^{g_1,...g_p} vol \equiv \sumong
\rho_{g_1,...g_p}(g)~\epsilon^{g_1,...g_p}
  \en
Due to the biinvariance of the volume form, the integral map $\int
: \Ga^{\we p} \mapsto \Cb$ satisfies the biinvariance conditions:
 \eq
  \int \Lcal_g f = \int f = \int \Rcal_g f
  \en

  Moreover, under the assumption that the volume form belongs to
  a nontrivial cohomology class, that is $d ( vol) = 0$ but
  $vol \not= d \rho$, the important property holds:
  \eq
  \int df =0
  \en
  with $f$  any $(p-1)$-form: $f=f_{g_2,...g_p}~ \theta^{g_2}
\we ... \we \theta^{g_p}$. This property, which allows
  integration by parts, has a simple proof. Rewrite
  $\int df$ as:
  \eq
\int df= \int (d f_{g_2,...g_p})\theta^{g_2} \we ... \we
\theta^{g_p}+ \int  f_{g_2,...g_p} d ( \theta^{g_2} \we ... \we
\theta^{g_p})
  \en
 Under the cohomology assumption the second term in the r.h.s.
 vanishes, since $d ( \theta^{g_2} \we ... \we
\theta^{g_p}) =0$ (otherwise, being a $p$-form, it should be
proportional to $vol$, and this would contradict the assumption
 $vol \not= d \rho$). Using now (\ref{partflat}) and
 (\ref{intpform}):
\eqa
 & &\int df= \int (t_{g_1} f_{g_2,...g_p})\theta^{g_1}\we \theta^{g_2} \we ...
\we \theta^{g_p} = \int
[\Rcal_{g_1^{-1}}f_{g_2,...g_p}-f_{g_2,...g_p}] \epsilon^{g_1,...g_p}
 vol = \nonumber \\
 & &~~~~~~~ = \epsilon^{g_1,...g_p}~
 \sumong [\Rcal_{g_1^{-1}}f_{g_2,...g_p}(g)-f_{g_2,...g_p}(g)]=0
\ena
 Q.E.D.

\sect{Bicovariant calculus on $S_3$}

In this Section we illustrate the general theory on
the particular example of the permutation group $S_3$.
 \sk
 Elements: $a=(12)$, $b=(23)$, $c=(13)$, $ab=(132)$, $ba=(123)$, $e$.
 \sk
 Multiplication table:
 \sk
 \begin{tabular}{|c|c|c|c|c|c|c|}
   \hline
     & e & a & b & c & ab & ba \\  \hline
   e & e & a & b & c & ab & ba \\   \hline
   a & a & e & ab & ba & b & c \\ \hline
   b & b & ba & e & ab & c & a \\ \hline
   c & c & ab & ba & e & a & b \\ \hline
   ab & ab & c & a & b & ba & e \\ \hline
   ba & ba & b & c & a & e & ab \\ \hline
 \end{tabular}
 \sk
 Nontrivial conjugation classes: $I = [a,b,c]$, $II = [ab,ba]$.
 \sk
There are 3 bicovariant calculi $BC_I$, $BC_{II}$, $BC_{I+II}$
corresponding to the possible unions of the conjugation classes
\cite{DMGcalculus}.
They have respectively dimension 3, 2 and 5. We examine here the
$BC_I$ and $BC_{II}$ calculi.

 \subsection{ $BC_I$ differential calculus}

Basis of the 3-dimensional vector space of one-forms:
 \eq
\theta^a,~\theta^b,~\theta^c
\en
\noi Basis of the 4-dimensional vector space of two-forms:
 \eq
  \theta^a \we \theta^b,~
 \theta^b \we \theta^c,~\theta^a \we \theta^c,~\theta^c \we \theta^b
 \en

 Every wedge product of two $\theta$ can be expressed
 as linear combination of the basis elements:
 \eq
  \theta^b \we \theta^a = -\theta^a \we \theta^c - \theta^c \we
  \theta^b,~~\theta^c \we \theta^a=-\theta^a \we \theta^b-\theta^b \we
  \theta^c
  \en

\noi Basis of the 3-dimensional vector space of three-forms:
 \eq
  \theta^a \we \theta^b \we \theta^c,~\theta^a \we \theta^c \we \theta^b,
  ~\theta^b \we \theta^a \we \theta^c
  \en

 and we have:
  \eqa
 & &\theta^c \we \theta^b \we \theta^a=- \theta^c \we \theta^a \we
\theta^c=-\theta^a \we \theta^c \we \theta^a=\theta^a \we \theta^b
\we \theta^c \nonumber \\
 & &\theta^b \we \theta^c \we \theta^a=- \theta^b \we \theta^a \we
\theta^b=-\theta^a \we \theta^b \we \theta^a=\theta^a \we \theta^c
\we \theta^b \nonumber \\
 & &\theta^c \we \theta^a \we \theta^b=- \theta^c \we
\theta^b \we \theta^c=-\theta^b \we \theta^c \we \theta^b=\theta^b
\we \theta^a \we \theta^c
\ena

\noi Basis of the 1-dimensional vector space of four-forms:

\eq
 vol = \theta^a \we \theta^b \we \theta^a \we \theta^c
 \en

 and we have:
  \eq
 \theta^{g_1} \we \theta^{g_2} \we \theta^{g_3} \we \theta^{g_4}=
 \epsilon^{g_1,g_2,g_3,g_4} vol \label{epsiI}
 \en
 where the nonvanishing components of the $\epsilon$ tensor are:
 \eqa
 & &
 \epsilon_{abac}=\epsilon_{acab}=\epsilon_{cbca}=\epsilon_{cacb}=
 \epsilon_{babc}=\epsilon_{bcba}=1 \\
 & &
 \epsilon_{baca}=\epsilon_{caba}=\epsilon_{abcb}=\epsilon_{cbab}=
 \epsilon_{acbc}=\epsilon_{bcac}=-1 \label{epsivaluesI}
 \ena
 Note the interesting property
 \eq
 f ~vol = vol~f,~~~\forall f \in Fun(G) \label{fvol}
 \en
 due to $\Rcal_a \Rcal_b \Rcal_a \Rcal_c = \Rcal_{caba}=\Rcal_e = id$
\sk
 \noi Cartan-Maurer equations:
\eqa
 & & d\theta^a+\theta^b\we\theta^c+\theta^c \we \theta^b =0
 \nonumber \\
 & & d\theta^b+\theta^a\we\theta^c+\theta^c \we \theta^a =0
 \nonumber \\
 & & d\theta^c+\theta^a\we\theta^b+\theta^b \we \theta^a =0
 \ena

 The exterior derivative on any three-form
of the type $\theta \we \theta \we \theta$ vanishes, as one can
easily check by using the Cartan-Maurer equations and the
equalities between exterior products given above. Then, as shown
in the previous Section,  integration of a total differential
vanishes on the ``group manifold" of $S_3$ corresponding to the
$BC_I$ bicovariant calculus. This ``group manifold" has three
independent directions, associated to the cotangent basis
$\theta^a,~\theta^b,~\theta^c$. Note however that the volume
element is of order four in the left-invariant one-forms $\theta$.
\sk
 \subsection{ $BC_{II}$ differential calculus}

Basis of the 2-dimensional vector space of one-forms:
 \eq
\theta^{ab},~\theta^{ba}
\en
\noi Basis of the 1-dimensional vector space of two-forms:
 \eq
  vol = \theta^{ab} \we \theta^{ba}=-\theta^{ba} \we \theta^{ab}
 \en
so that:
  \eq
 \theta^{g_1} \we \theta^{g_2}=
 \epsilon^{g_1,g_2} vol \label{epsiII}
 \en
 where the $\epsilon$ tensor is the usual 2-dimensional Levi-Civita tensor.
 Again $f~vol = vol~f$ since $abba=e$.
 \sk
\noi Cartan-Maurer equations:
\eq
 d\theta^{ab} =0,~~ d\theta^{ba} =0
 \en

 Thus the exterior derivative on any one-form $\theta^g$
 vanishes and integration of a total differential
vanishes on the group manifold of $S_3$ corresponding to the
$BC_{II}$ bicovariant calculus. This group manifold has two
independent directions, associated to the cotangent basis
$\theta^{ab},~\theta^{ba}$.

\subsection{Visualization of the $S_3$ group ``manifold"}

 We can draw a picture of the group manifold of $S_3$. It is made
 out of 6 points, corresponding to the group elements and identified with
 the functions $x^e,x^a,x^b,x^c,x^{ab},x^{ba}$.
\sk
$BC_I$ - calculus:
\sk
 From each
 of the six points $x^g$ one can move in three directions, associated to
 the tangent vectors $t_a,t_b,t_c$, reaching three other points
 whose ``coordinates" are
 \eq
\Rcal_a x^g = x^{ga},~~\Rcal_b x^g = x^{gb},~~\Rcal_c x^g = x^{gc}
\en
 The 6 points and the ``moves" along the 3 directions are
 illustrated in the Fig. 1. The links are not oriented since
 the three group elements $a,b,c$ coincide with their inverses.

 \sk
$BC_{II}$ - calculus:
\sk
 From each
 of the six points $x^g$ one can move in two directions, associated to
 the tangent vectors $t_{ab},t_{ba}$, reaching two other points
 whose ``coordinates" are
 \eq
\Rcal_{ab} x^g = x^{gba},~~\Rcal_{ba} x^g = x^{gab}
\en
 The 6 points and the ``moves" along the 3 directions are
 illustrated in Fig. 1. The arrow convention on a link
 labeled  (in italic) by a group element $h$
 is as follows: one
 moves in the direction of the arrow via the action
 of $\Rcal_{h}$ on $x^g$. (In this case $h=ab$). To move in the opposite
 direction just take the inverse of $h$.
\sk

 The pictures in Fig. 1
 characterize the bicovariant calculi  $BC_I$ and $BC_{II}$ on $S_3$,
 and were drawn in  Ref. \cite{DMGcalculus} as examples of
 digraphs, used to characterize different calculi on sets.
 Here we emphasize  their geometrical meaning as
  finite group ``manifolds".

%\sk \iffigs
%\begin{figure}[h]
%\epsfxsize = 10cm \centerline{\epsfig{figure=S3.eps}}
%\caption{$S_3$ group manifold, and moves of the points under the
%group action}
%\label{s3}
% \unitlength=1mm
%\end{figure}
%\fi

%%Begin InstantTeX Picture
\let\picnaturalsize=N
\def\picsize{4.0in}
\def\picfilename{S3.eps}
%If you do not have the picture file add:
%\let\nopictures=Y
%to the beginning of the file.
\ifx\nopictures Y\else{\ifx\epsfloaded Y\else\input epsf \fi
\let\epsfloaded=Y
\centerline{\ifx\picnaturalsize N\epsfxsize \picsize\fi
\epsfbox{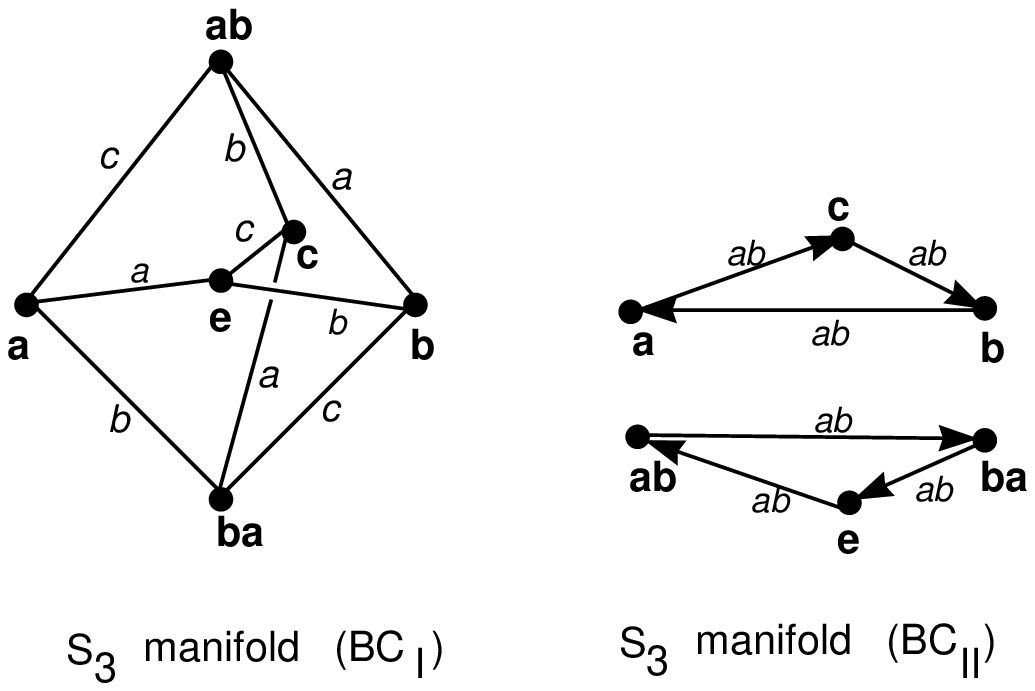}}}\fi
%%End InstantTeX Picture
\sk
  {\small{\bf Fig. 1} : $S_3$ group manifold, and moves of the
   points under the group action}
 \sk
 \subsection{Distance matrix}

We define a distance matrix $(dist)_{i,j}$ between two points
$i,j$ of the finite group manifold as the minimum number of links
that connects them. It is easy to verify that the graphs in Fig. 1
(and more generally any graph corresponding to a bicovariant
calculus on a finite $G$)  are $ad(G)$ invariant, and therefore
$dist$ itself is $ad(G)$ invariant and can be taken as biinvariant
metric. In the case of the connected manifold corresponding to
$BC_I$ the distance matrix is invertible and given by:

\eq
 dist=\left( \matrix{ 0 & 1 & 1 & 1 & 2 & 2 \cr 1 & 0 & 2
 & 2 & 1 & 1 \cr 1 & 2 & 0 & 2 & 1 & 1
   \cr  1 & 2 & 2 & 0 & 1 & 1 \cr  2 & 1 & 1
   & 1 & 0 & 2 \cr 2 & 1 & 1 & 1 & 2 & 0
   \cr  }\right)
   \en
where rows and columns are ordered as $e,a,b,c,ab,ba$.

For the disconnected $BC_{II}$ graph we must define also the
distance between two disconnected points, and we arbitrarily set
it to zero. The resulting distance matrix is also invertible.

\sect{Softening $G$ and ``gravity" on $S_3$}

We have in mind to construct a dynamical theory of vielbein fields
whose vacuum solution describes the $G$ manifold. In particular
let us take $G = S_3$ in the $BC_I$ setting. Then the dynamical
fields of the theory are collected in the 1-form vielbein $V^g$,
which is not left-invariant any more since it is a
deformation of $\theta^g$:
 \eq
  V^a = \sumongp \V{a}{g} (x) \theta^g,~~V^b =\sumongp \V{b}{g} (x)
  \theta^g,~~V^c = \sumongp \V{c}{g} (x) \theta^g
  \en
  In addition we consider also the ``spin connection"
  1-form $\omc{g_1}{g_2}$ as an independent field
  (first order formulation). The field equations will
  determine the expression  of $\om$ in terms
  of the vielbein field.

We try then to mimic the Einstein-Cartan action of general
relativity:
 \eq
   A=\int \V{g_3}{h_3} \V{g_4}{h_4} \epsilon_{g_1,g,g_3,g_4} \ga^{g,g_2}
   \R{g_1}{g_2} \we \theta^{h_3} \we
   \theta^{h_4} \label{ECaction}
   \en
where $\R{g_1}{g_2}$, the ``soft" curvature, is given in terms of
$\om$ as in eq. (\ref{curvature}) and the indices of the $\epsilon$
tensor (defined in (\ref{epsiI})) are lowered with the (bi-invariant)
metric
$\ga_{g_1,g_2}=\de_{g_1,g_2}$. The integrand is a 4-form (since we
chose the $BC_I$ calculus on $S_3$): thus the action is formally
identical to the one of general relativity. The one-forms $\theta$
are the vielbeins of $S_3$ discussed in the previous Section: they
are given one-form fields without dynamics.
 \sk
 {\bf Invariances of A}
  \sk
  Consider the field transformations:
\eqa
  & &(\V{g'}{h})'=\a{g'}{g} \V{g}{h} \\
  & &(\omc{g'}{h'})'=\a{g'}{g} \omc{g}{h} \ainv{h}{h'} + \a{g'}{h} d
  \ainv{h}{h'} \label{atrans}
  \ena
Requiring $A$ to be invariant under these transformations sets
some conditions on the $x$ dependent ``rotation" matrix $a$.
Recalling that the curvature transforms as $(\R{g'}{h'})'=
\a{g'}{g} \R{g}{h} \ainv{h}{h'}$ we find the transformed action
$A'$:
 \eq
 A'=\int \V{k_3}{h_3} \V{k_4}{h_4}\epsilon_{g_1,g,g_3,g_4} \ga^{g,g_2}
  \a{g_3}{k_3} \a{g_4}{k_4} \a{g_1}{k_1}
  \R{k_1}{k_2} \ainv{k_2}{g_2} \we \theta^{h_3} \we
   \theta^{h_4}
   \en
In the case of usual general relativity, the Lorenz metric and the
Levi-Civita tensor are conserved under local Lorenz rotations, and
this implies the invariance of the action under local Lorenz
transformations.

Here the $\epsilon$ tensor is the one given in
(\ref{epsivaluesI}); moreover the $x$ dependent $a$ matrix
elements do not commute with the two-form $\R{k_1}{k_2}$ as in the
usual case.

We will show that if the $a$ matrices entries are taken to be the
 functions of eq. (\ref{Gadjoint})  satisfying an additional
periodic condition, then the action is invariant under the
transformations (\ref{atrans}).

First, for the adjoint matrices of (\ref{Gadjoint}) we have
$\ga^{g,g_2} \ainv{k_2}{g_2}=\a{g}{g_2} \ga^{g_2,k_2}$ because of
(\ref{ginv}). Suppose then that the $a$ matrix entries satisfy a
``two unequal links" periodic condition:
 \eq
   \a{i}{j}=\Rcal_{gg'} \a{i}{j},~~~g \not= g' \label{twolinks}
   \en
   Then we can bring the $\a{g}{g_2}$ term to the left of the
   curvature two-form (the $g \not= g'$ in (\ref{twolinks})
   is due to  $\theta^g \we \theta^g=0$), and we
 see that the action $A$ is invariant if:
  \eq
  \epsilon_{g_1,g_2,g_3,g_4} \a{g_1}{k_1} \a{g_2}{k_2}
  \a{g_3}{k_3} \a{g_4}{k_4} = \epsilon_{k_1,k_2,k_3,k_4}
  \label{epsiloncons}
  \en
i.e. if $\epsilon$ is a conserved tensor. But as we have already
argued in Section 2.11 this is the case when the $a$ matrix rotates the indices
according to the adjoint action of $G$.

Hence we have an action invariant under the local $ad(G)$
transformations, in analogy with the local Lorenz rotations of
general relativity.

This action is also invariant under the analogue of general
coordinate transformations. Indeed diffeomorphisms along a generic
tangent vector $v$ are  generated by the Lie derivative
 $l_v = d i_v + i_v d$. Then under diffeomorphisms the variation
 of $A$ is given by
 \eq
 \de A=\int l_v (\mbox{4-form}) = \int [d i_v \mbox{(4-form)}
  + i_v d \mbox{(4-form)}]=0 \label{lieA}
\en
since $d\mbox{ (4-form)}=0$ and $\int d\mbox{ (3-form)} =0$. \sk
{\bf Field equations} \sk The field equations are obtained by
varying the action with respect to the dynamical fields
$\V{g}{h}$, $\omc{g}{h}$. The $\de \omc{g}{h}$ variation yields an
equation relating $\omc{g}{h}$ to (first derivatives of) the
vielbein and its inverse, as in the usual zero-torsion condition
of ordinary Einstein-Cartan gravity. Varying with respect to
$\V{g}{h}$ leads to the analogues of Einstein eqs.:
 \eq
  \V{g_3}{h_3}  \ga^{g_0,g_2}
   \R{g_1}{g_2,h_1,h_2} [\epsilon_{g_1,g_0,g,g_4} \epsilon^{h_1,h_2,h,h_4}
+\epsilon_{g_1,g_0,g_3,g} \epsilon^{h_1,h_2,h_3,h}]=0
\en
where the curvature components $\R{g_1}{g_2,h_1,h_2}$ are defined
by $\R{g_1}{g_2}=\R{g_1}{g_2,h_1,h_2} \theta^{h_1} \we
\theta^{h_2}$.
 \sk
{\bf Note 4.1} : the analogue of a cosmological term
 \eq
 \int \epsi_{g_1,g_2,g_3,g_4} \V{g_1}{h_1} \V{g_2}{h_2}
 \V{g_3}{h_3}\V{g_4}{h_4} ~\theta^{h_1} \we \theta^{h_2} \we
 \theta^{h_3} \we \theta^{h_4}
 \en
is invariant under the local tangent rotations (\ref{atrans})
because of property (\ref{epsiloncons}), and under diffeomorphisms
because of (\ref{lieA}. Adding this term to the action
(\ref{ECaction}) allows ``vacuum" solutions with $\R{g_1}{g_2,h_1,h_2}
\not= 0$.
\sk

{\bf Note 4.2} :  The same action (\ref{ECaction}) can be used in the
case of the finite group $Z^N \times Z^N \times Z^N \times Z^N $.
Here the situation simplifies: for example the $\epsilon$ tensor
becomes the usual Levi-Civita tensor, the basic one forms
anticommute etc. One then obtains a discretized gravity of the
type discussed in ref. \cite{DMgrav}, with some differences. The field
equations are derived from a variational principle, the local
symmetry involves functions arbitrary up to a (``two unequal
links") periodicity condition, no procedure is used to ``localize"
the components of the curvature tensor, and no use of the left
symmetric tensor product $\otimes_L$ is made.

\sk
{\bf Note 4.3}: the analogue of the topological action:
\eq
\int \R{g_1}{g_2} \we \R{g_2}{g_1}
\en
has all the invariances discussed above, even without
the two-links periodic condition on the matrices
$a(x)$, thanks to $[vol, f]=0$.

 \sk\sk\sk
 {\bf Acknowledgments}
 \sk
 \noi I have benefited from discussions with P. Aschieri and  F.
 Mueller-Hoissen.
\sk\sk

\app{Lie derivative along generic tangent fields}

As discussed in ref. \cite{Athesis} a generic vector field $V$ can be
written in terms of the left-invariant tangent vectors $t_i$ of
eq. (\ref{tg}) as $V=a^i t_i$, $a^i \in Fun(G)$ and defined to act
on any $b \in G$ as:
 \eq
 V b=(a^i t_i) b \equiv a^i (t_i b)
 \en
We denote by $\Xi$  the space of vector fields $V$. The product
between elements of $Fun(G)$ and left-invariant tangent vectors
$t_i$ generalizes to the whole $\Xi$: $(aV)b \equiv a(Vb)$, and
 \eq
  (a+b)V= aV+ bV;~~(ab)V=
 a(bV) ;~~ (\lambda a )V= \lambda (aV)
 ,~~\lambda \in \Cb
 \en
 $\Xi$ is the analogue of the space of derivations on $Fun(G)$.
 Indeed:
 \eqa
 & & V(a+b)=V(a)+V(b),~~V(\la a)=\la V(a)~~~~~~~\mbox{Linearity}\\
 & & V(ab)\equiv (c^i t_i )(ab)=t_i(a)(\Rcal_{i^{-1}}
 b)c^i + a V(b)~~~\mbox{Leibniz~rule}
 \ena

{\bf Inner derivative}
 \sk
 The $i_V$ contraction operator  is defined by the following
 properties ($V \equiv c^i t_i$, $a \in Fun(G)$, $\la \in \Cb$):
 \eqa
 & &i_V(\rho) =i_{t_i}(\rho) c^i\\
 & &i_V(a)=0\\
 & &i_V(\theta^i)=c^i\\
 & &i_V(\rho \we \rhop)=i_{t_i}(\rho) \we (\Rcal_{i^{-1}} \rhop)
 c^i + (-1)^{deg(\rho)} \rho \we i_V(\rhop)\\
 & &i_V(a\rho + \rhop)=ai_V(\rho) + i_V(\rhop)\\
 & &i_V(\rho a)=i_{t_i}(\rho)(\Rcal_{i^{-1}} a) c^i \\
 & &i_{\la V}=\la i_V
 \ena
\sk\sk
 {\bf Lie derivative}
 \sk
 Definition:
 \eq
 l_V \equiv i_V d + d i_V
 \en
 Properties ($V \equiv c^i t_i$, $a,b,c^i \in Fun(G)$):
 \eqa
  & & l_V(a)=V(a) \\
  & & l_V d\rho=d l_V \rho\\
  & & l_V(\la \rho+\rhop)=\la l_V(\rho)+l_V(\rhop)\\
  & & l_{bV}(\rho)= (l_V\rho)b-(-1)^{deg(\rho)}i_V(\rho) \we db\\
  & & l_V (\rho\we\rhop)=\rho \we l_V (\rhop) + l_{t_i}(\rho)
  \we (\Rcal_{i^{-1}} \rhop) c^i + \nonumber \\
  & & ~~~~~~~+(-1)^{deg(\rhop)} i_{t_i}(\rho)
  \we (\Rcal_{i^{-1}} \rhop) dc^i\\
  & & l_{t_i}(\theta^j)=\C{j}{k,i} \theta^k
  \ena

The proof that the Lie derivative $l_{t_i}$ defined as $l_{t_i}
\rho=(id \otimes t_i) \DR (\rho)$ in (\ref{liederal}) is equal to
$ l_{t_i}= i_{t_i} d + d i_{t_i}$ is done by induction on the
generic $p$-form $a_{i_1...i_p} \theta^{i_1}\we...\we
\theta^{i_p}$, checking first that both definitions give the same
result on $\theta^j$ (see ref.s (\cite{ACintro,Athesis})).

%%%%%%%%%%%%%%%%%%%%%%%%%%%%%%%%%%%

\vfill\eject
\end{document}